# Failures of "meso-phase" hypothesis near vapor-liquid critical point

I.H. Umirzakov

*Institute of Thermophysics, Novosibirsk, Russia, e-mail: cluster125@gmail.com*

**Abstract**

It is shown that the "meso-phase" hypothesis of Woodcock L. V. fails to describe quantitatively and qualitatively the isochoric and isobaric heat capacities, speed of sound, long wavelength limit of the structural factor, isothermal compressibility, density fluctuations, Joule-Thompson coefficient and isothermal throttling coefficient of argon in the "meso-phase" region. It is also shown that VdW-EOS can describe qualitatively the excess Gibbs energy and rigidity of argon near critical point.

## 1. Introduction

There are two alternative theories of critical region of liquid-vapor first order phase transitions. The first theory is the traditional theory of the region near single critical point based upon Ising-like scaling theory with crossover to classical equations of state [1-25]. VdW-EOS [6] and the fundamental equations of state [17-20], which are based upon the concept of a single gas-liquid critical point, are representations of these classical equations of state.

The second theory is the "meso-phase" hypothesis of Woodcock [26]. According to the "meso-phase" hypothesis at critical and supercritical temperatures on the thermodynamic (density, pressure)-plane exists a region where the pure substance is in "meso-phase", which consists of small clusters that are gas like and clusters of macroscopic size that are liquid like, there is exist a line of critical points over a finite range of densities at critical temperature and pressure instead of single critical point, and the pressure in the "meso-phase" is linear function of density. This hypothesis is reminiscent of an old concept of the supercritical fluid as a mixture of "gasons" and "liquidons" that has turned out to be inconsistent with the experimental evidence [4,5].

Some predictions of the "meso-phase" hypothesis were criticized by Sengers and Anisimov [4] and Umirzakov [27]. According to [4] in contrast to the conjecture of Woodcock, there is no reliable experimental evidence to doubt the existence of a single critical point in the thermodynamic limit and of the validity of the scaling theory for critical thermodynamic behavior.

According to [26] the Van der Waals critical point does not comply with the Gibbs phase rule and its existence is based upon a hypothesis rather than a thermodynamic definition. The paper [27] mathematically demonstrates that a critical point is not only based on a hypothesis that is used to define values of two parameters of VdW-EOS. Instead, the author of [27] argued that a critical point is a direct consequence of the thermodynamic phase equilibrium conditions resulting in a single critical point. It was shown that the thermodynamic conditions result in the first and second partial derivatives of pressure with respect to volume at constant temperature at a critical point equal to zero which are usual conditions of an existence of a critical point [27].

The papers [28] and [29] were the responses to the critique of some predictions of the hypothesis in [4] and [27], respectively.

The paper [28] was criticized in [30]. It was shown [30] that: (1) the expressions for the isochoric and isobaric ($C_P$) heat capacities of liquid and gas, coexisting in phase equilibrium, the heat capacities at saturation of liquid and gas ($C_\sigma$) and the heat capacity ($C_\lambda$) used in Woodcock's article [28] are incorrect; (2) the conclusions of the article based on the comparison of the incorrect $C_V$, $C_P$, $C_\sigma$ and $C_\lambda$ with experimental data are also incorrect; (3) the lever rule used in [28] cannot be used to define $C_V$ and $C_P$ in the two-phase coexistence region; (4) a correct expression for the isochoric heat capacity describes the experimental data well; (5) there is no misinterpretation of near-critical gas–liquid heat capacity measurements in the two-phase coexistence region; (6) there are no proofs in the article that: (a) the divergence of $C_V$ is apparent; (b) it has not been established experimentally that the thermodynamic properties of fluids satisfy scaling laws with universal critical exponents asymptotically close to a single critical point of the vapor–liquid phase transition; and (c) there is no singular critical point on Gibbs density surface. Many mathematical and logical errors were also found in [28].

As known the theory is wrong if it does not agree with experiment.

It has been established experimentally that the thermodynamic properties of fluids satisfy scaling laws with universal critical exponents asymptotically close to a single critical point of the vapor–liquid phase transition [1,4]. The fundamental equations of state [17-20] represent the available experimental data on the (pressure, temperature, density)-relation, liquid-vapor coexistence, isochoric and isobaric heat capacities, speed of sound, Joule-Thomson coefficient, isothermal throttling coefficient and etc., typically within experimental uncertainty. So the scaling theory and FEOS are in good agreement with the available experimental data.

The comparison of the predictions of the "meso-phase" hypothesis with the experimental data can show that the "meso-phase" hypothesis is wrong or not.

The fundamental equation of state of argon of Tegeler-Span-Wagner (TSW-EOS) [17] represents the available data for the accurate (pressure, temperature, density)-data [21], data on the liquid-vapor coexistence [22], available data on the isochoric and isobaric heat capacities, speed of sound, Joule-Thomson coefficient, isothermal throttling coefficient and etc., typically within experimental uncertainty as required by fundamental thermodynamic relationships. Therefore we compare the predictions of the hypothesis for the pressure, isothermal rigidity coefficient, isochoric and isobaric heat capacities, speed of sound, isothermal compressibility, density fluctuations, optic (long wavelength) limit of the structural factor, Helmholtz energy, Gibbs energy, entropy, internal energy, enthalpy, Joule-Thomson coefficient and isothermal throttling coefficient of argon with the corresponding predictions of TSW-EOS in Chapter **2**.

Chapter **3** contains the response to the critique by Woodcock [29] of VdW-EOS, the parametric solution of the equations of liquid-vapor phase equilibrium of Van der Waals fluid (VdW-fluid) and some assertions of the paper [27].

**Appendix** shows that the paper [29] includes many incorrect equations, mathematical and logical errors.

## 2. Comparison of predictions of "meso-phase" hypothesis with available data for argon

The predictions of the "meso-phase" hypothesis for the isochoric and isobaric heat capacities, speed of sound, isothermal compressibility, density fluctuations, optic (long wavelength) limit of the structural factor, Helmholtz energy, Gibbs energy, entropy, internal energy, enthalpy, Joule-Thomson coefficient and isothermal throttling coefficient of argon are obtained and discussed in Chapter **2.1**. The expressions to define the above physical properties of argon using TSW-EOS [17] are given in Chapter **2.2**. The comparison of the predictions of the "meso-phase" hypothesis and TSW-EOS for argon is presented and discussed in Chapter **2.3**.

### 2.1. Predictions of the "meso-phase" hypothesis for argon

According to [26] pressure $p_M(\rho,T)$ of the "meso-phase", which exists in the density interval $\rho_B(T) \leq \rho \leq \rho_A(T)$, is a linear function of density (further the subscript "$M$" of a quantity means that the quantity is obtained according to the "meso-phase" hypothesis):

$$p_M(\rho,T) = p_0(T) + \omega\rho, \tag{1}$$

where $T$ is temperature, $\rho$ is the mass density and $\omega = (\partial p/\partial \rho)_T$ is the isothermal rigidity coefficient, which does not depend on density, so $\omega = \omega_M(T)$. The thermodynamic relation $(\partial C_V/\partial v)_T = T(\partial^2 p/\partial T^2)_V$ [8], where $v = m/\rho$ and $m$ is the mass of the particle (atom or molecule), and Eq. 1 give

$$C_{VM}(\rho,T) = C_{VM}(\rho_B,T) + Tm \cdot d^2 p_0/dT^2 \cdot (1/\rho - 1/\rho_B) - Tm \cdot d^2\omega_M/dT^2 \cdot \ln(\rho/\rho_B). \tag{2}$$

We conclude from Eq. 2 that in the case when the equalities $d^2 p_0/dT^2 = 0$ and $d^2\omega_M/dT^2 = 0$ are valid in the interval $\rho_B \leq \rho \leq \rho_A$ the equality $C_{VM}(\rho,T) = C_{VM}(\rho_B,T)$ takes place in this interval. Therefore $C_{VM}(\rho,T)$ is independent of density in this interval for this case because $C_{VM}(\rho_B,T)$ does not depend on density. One can also conclude from Eq. 2 that in the cases when 1) $d^2 p_0/dT^2 \neq 0$ or 2) $d^2\omega_M/dT^2 \neq 0$ or 3) $d^2 p_0/dT^2 \neq 0$ and $d^2\omega_M/dT^2 \neq 0$ in this interval the isochoric heat capacity $C_{VM}(\rho,T)$ is nonlinear function of density. So $C_{VM}$ is equal to constant or it is a nonlinear function of density in the density interval $\rho_B \leq \rho \leq \rho_A$ in general case.

According to Figs. 4 and 5 [26] and Table 1 [26] the relations

$$T = a_1 \omega_M(T) + T_{cM}, \tag{3a}$$

$$p_0(T) = p_{cM} - b_1 \cdot (T - T_{cM}), \tag{3b}$$

where $T_{cM} = 151.2136\,K$, $p_{cM} = 4.9493\,MPa$, $a_1 = 2239.5\,m^{-3}kg \cdot MPa^{-1}K$ and $b_1 = 0.04168\,K^{-1}MPa$, are valid for argon in the interval $153\,K \leq T \leq 157\,K$. The above value of critical temperature $T_{cM}$ was obtained in [26] by extrapolation of Eq. 3a to the region $T < 153\,K$. The above value of critical pressure $p_{cM}$ was evaluated in [26] by extrapolation of Eq. 3b to region $T < 153\,K$ and using above value of $T_{cM}$. The above extrapolations have no a physical or theoretical basis.

We conclude from Eqs. 2, 3a and 3b that

$$C_{VM}(\rho,T)/k = c \tag{4}$$

in interval $\rho_B \leq \rho \leq \rho_A$, because $d^2 p_0/dT^2 = 0$ and $d^2\omega_M/dT^2 = 0$. Here $c = c(T) \equiv C_{VM}(\rho_B,T)/k$ and $k$ is the Boltzmann's constant, $k = 1.380648 \cdot 10^{-23}\,K^{-1}J$. The values of $\rho_A$ and $\rho_B$ for argon are given in table 1 [26].

The following thermodynamic relations [7,8]

$$C_P(\rho,T) = C_V(\rho,T) + mT(\partial p/\partial T)_\rho^2/\rho^2(\partial p/\partial \rho)_T, \tag{5}$$

$$c_s^2(\rho,T) = (\partial p/\partial \rho)_T + mT \cdot (\partial p/\partial T)_\rho^2/\rho^2 C_V(\rho,T) \tag{6}$$

and Eqs. 1-6 give the following relations

$$p_M(\rho,T)m/\rho kT = t\beta\gamma/r + \gamma(1-\lambda/r)(1-t), \tag{7}$$

$$\omega_M(T)m/kT = \gamma(1-t), \tag{8}$$

$$C_{PM}(\rho,T)/k = c + \gamma(1-\lambda/r)^2/(1-t), \tag{9}$$

$$mc_{sM}^2(\rho,T)/kT = \gamma(1-t) + \gamma^2(1-\lambda/r)^2/c, \tag{10}$$

where $t = T_{cM}/T$, $r = \rho/\rho_{Bc}$, $\rho_{Bc} \equiv \rho_B(T_{cM}) = 475.61\, kg \cdot m^{-3}$, $\lambda = a_1 b_1/\rho_{Bc} = 0.196$, $\beta = a_1 p_{cM}/\rho_{Bc} T_{cM} = 0.154$, $m = 6.634 \cdot 10^{-26}\, kg$ is the mass of the atom of argon, $\gamma = m/ka_1 = 2.146$, $C_{PM}(\rho,T)$ is the isobaric heat capacity and $c_{sM}(T,\rho)$ is the speed of sound of argon.

We have from Eqs. 9-10

$$(\partial C_{PM}/\partial\rho)_T = 2mT(1 - a_1 b_1/\rho)b_1/(T - T_{cM})\rho^2, \tag{11}$$

$$(\partial c_{sM}^2/\partial\rho)_T = 2mT(1 - a_1 b_1/\rho)b_1/a_1 c\rho^2. \tag{12}$$

Using Eqs. 11-12 and taking into account the inequality $\rho_B > a_1 b_1 = 93.34\, m^{-3} kg$ (see Table 1 [26]) we conclude that the isobaric heat capacity and the speed of sound increase with increasing density in the interval $\rho_B \leq \rho \leq \rho_A$ for $153\, K \leq T \leq 157\, K$. We have from Eqs. 11-12

$$(\partial C_{PM}/\partial\rho)_T/(\partial c_{sM}^2/\partial\rho)_T = a_1 c/(T - T_{cM}). \tag{13}$$

Therefore the ratio of the slopes of the isotherms of the isobaric heat capacity and speed of sound does not depend on the density.

We obtain from Eq. 7

$$\chi_{TM}(\rho,T)\rho kT/m = \gamma^{-1}(1-t)^{-1}. \tag{14}$$

for the isothermal compressibility $\chi_T \equiv 1/\rho(\partial p/\partial\rho)_T$ [11]. One can see from Eq. 14 that the isothermal compressibility is inversely proportional to density. We conclude from Eqs. 9, 10 and 14 that $C_{PM}$ and $c_{sM}^2$ are the quadratic functions of volume $v$, and $\chi_{TM}$ is the linear function of volume.

The relation $\Delta = kT/m(\partial p/\partial\rho)_T\big|_{\rho=<\rho>\equiv m<N>/V}$ is valid for the quantity $\Delta \equiv [<N^2> - <N>^2]/<N>$, which characterizes the fluctuations of number of particles in the macroscopic container having a volume $V$ and walls penetrable for the particles [11]. The standard deviation of fluctuations of the density $\rho = mN/V$ is defined by $<\rho^2> - <\rho>^2 = <\rho>\cdot\Delta/V$, so $\Delta$ also characterizes the density fluctuations [8]. Here $<N>$ is the mean of $N$ over its fluctuations and $k$ is the Boltzmann's constant. We have

$$\Delta_M(T) = \gamma^{-1}(1-t)^{-1}, \tag{15}$$

According to [11] the long wavelength limit of the structural factor $s_0$ is defined from $s_0 = kT\rho\chi_T/m$. Therefore we have using Eq. 7

$$s_{0M}(T) = \gamma^{-1}(1-t)^{-1}. \tag{16}$$

One can see from Eqs. 15-16 that $\Delta_M$ and $s_{0M}$ do not depend on density.

We obtain from Eq. 7, the exact relations $p(T,\rho) = -[\partial A(\rho,T)/\partial(m/\rho)]_T$, $S = -(\partial A/\partial T)_\rho$, $G = A + pm/\rho$, $E = A + TS$ and $H = G + TS$ [8], where $A$, $G$, $S$, $E$ and $H$ are the Helmholtz energy, Gibbs energy (chemical potential), entropy, internal energy and enthalpy per particle, respectively, the following relations ($r_B = \rho_B/\rho_{Bc}$)

$$A_M(\rho,T)/kT - A_{MB}/kT = \gamma[\beta t - \lambda(1-t)](1 - r_B/r)/r_B + \gamma(1-t)\ln(r/r_B), \tag{17}$$

$$S_M(\rho,T)/k - S_{MB}/k = \lambda\gamma(1 - r_B/r)/r_B - \gamma\ln(r/r_B), \tag{18}$$

$$E_M(\rho,T)/kT - E_{MB}/kT = t\gamma(\beta + \lambda)(1 - r_B/r)/r_B - t\gamma\ln(r/r_B), \tag{19}$$

$$G_M(\rho,T)/kT - G_{MB}/kT = (1-\gamma)\lambda(1-t)(1 - r_B/r)/r_B + \gamma(1-t)\ln(r/r_B), \tag{20}$$

$$H_M(\rho,T)/kT - H_{MB}/kT = \lambda\gamma(1 - r_B/r)/r_B - t\gamma\ln(r/r_B), \tag{21}$$

$A_{MB} = A_M(\rho_B,T)$, $S_{MB} = S_M(\rho_B,T)$, $E_{MB} = E_M(\rho_B,T)$, $G_{MB} = G_M(\rho_B,T)$, $H_{MB} = H_M(\rho_B,T)$.

We have from Eqs. 17-21 the inequalities $(\partial A_M/\partial\rho)_T > 0$, $(\partial G_M/\partial\rho)_T > 0$ and $(\partial S_M/\partial\rho)_T < 0$ because $T > T_{cM}$, $p_{cM} > b_1(T - T_{cM})$ and $\rho > a_1 b_1$ [26]. Therefore $A_M$ and $G_M$ increase and $S_M$ decreases with increasing density in the interval $\rho_B \leq \rho \leq \rho_A$.

We have from Eq. 7

$$v = m/[(p - p_{cM})/(T - T_{cM}) + b_1]a_1. \tag{22}$$

Using Eqs. 9 and 22 one can obtain

$$\mu_M(\rho,T)k\rho/m = \frac{t - \lambda/r}{c(1-t) + \gamma(1 - \lambda/r)^2} \tag{23}$$

from the relation $\mu \equiv (\partial T/\partial p)_H = [T(\partial v/\partial T)_P - v]/C_P$ for the Joule-Thomson coefficient [8].

We have from Eqs. 21-22

$$H_M = H_{MB} + Tb_1(m/\rho_B - m/a_1[(p - p_{cM})/(T - T_{cM}) + b_1])$$
$$- T_{cM}m/a_1 \cdot \ln(a_1[(p - p_{cM})/(T - T_{cM}) + b_1]/\rho_B). \tag{24}$$

We obtain from Eq. 24

$$\delta_{TM}(\rho,T)\rho/m = (\lambda/r - t)/(1-t) \tag{25}$$

for the isothermal throttling coefficient $\delta_T \equiv [\partial H(p,T)/\partial p]_T$ [17]. Eq. 25 shows that $\delta_{TM}$ is the linear function of volume. We can conclude from Eqs. 23 and 25 that $\mu_M = \delta_{TM} = 0$ if $T/T_{cM} = \rho/a_1 b_1$, $\mu_M > 0$ and $\delta_{TM} < 0$ if $T/T_{cM} < \rho/a_1 b_1$, and $\mu_M < 0$ and $\delta_{TM} > 0$ if $T/T_{cM} > \rho/a_1 b_1$, because $T/T_{cM} > 1$ and $c > 0$.

Eqs. 3a, 3b, 4 and 7-25 are valid in the interval $\rho_B(T) \leq \rho \leq \rho_A(T)$ for $153\ K \leq T \leq 157\ K$.

## 2.2. Predictions of the fundamental equation of state of argon

We have calculated the reduced pressure, isothermal rigidity coefficient, isochoric and isobaric heat capacities, speed of sound, isothermal compressibility, density fluctuations, optic (long wavelength) limit of the structural factor, Helmholtz energy, Gibbs energy, entropy, internal energy, enthalpy, Joule-Thomson coefficient and isothermal throttling coefficient of argon from the following thermodynamic relations (see Table 26 [17])

$$p(\rho,T)m/\rho kT = 1+\delta\alpha_\delta^r, \tag{26}$$

$$\omega(\rho,T)m/kT = 1+2\delta\alpha_\delta^r + \delta^2\alpha_{\delta\delta}^r, \tag{27}$$

$$C_V(\rho,T)/k = -\tau^2\alpha_{\tau\tau}, \tag{28}$$

$$C_P(\rho,T)/k = -\tau^2\alpha_{\tau\tau} + (1+\delta\alpha_\delta^r - \delta\tau\alpha_{\delta\tau}^r)^2/(1+2\delta\alpha_\delta^r + \delta^2\alpha_{\delta\delta}^r), \tag{29}$$

$$c_s^2(\rho,T)m/kT = 1+2\delta\alpha_\delta^r + \delta^2\alpha_{\delta\delta}^r - (1+\delta\alpha_\delta^r - \delta\tau\alpha_{\delta\tau}^r)^2/\tau^2\alpha_{\tau\tau}, \tag{30}$$

$$\chi_T(\rho,T)\rho kT/m = (1+2\delta\alpha_\delta^r + \delta^2\alpha_{\delta\delta}^r)^{-1}, \tag{31}$$

$$\Delta(\rho,T) = (1+2\delta\alpha_\delta^r + \delta^2\alpha_{\delta\delta}^r)^{-1}, \tag{32}$$

$$s_0(\rho,T) = (1+2\delta\alpha_\delta^r + \delta^2\alpha_{\delta\delta}^r)^{-1}, \tag{33}$$

$$A(\rho,T)/kT = \alpha, \tag{34}$$

$$G(\rho,T)/kT = 1+\alpha+\delta\alpha_\delta^r, \tag{35}$$

$$S(\rho,T)/k = \tau\alpha_\tau - \alpha, \tag{36}$$

$$E(\rho,T)/kT = \tau\alpha_\tau, \tag{37}$$

$$H(\rho,T)/kT = 1+\tau\alpha_\tau + \delta\alpha_\delta^r, \tag{38}$$

$$\mu(\rho,T)k\rho/m = -(\delta\alpha_\delta^r + \delta\tau\alpha_{\delta\tau}^r + \delta^2\alpha_{\delta\delta}^r)/[(1+\delta\alpha_\delta^r - \delta\tau\alpha_{\delta\tau}^r)^2 - \tau^2\alpha_{\tau\tau}(1+2\delta\alpha_\delta^r + \delta^2\alpha_{\delta\delta}^r)], \tag{39}$$

$$\delta_T(\rho,T)\rho/m = 1 - (1+\delta\alpha_\delta^r - \delta\tau\alpha_{\delta\tau}^r)/(1+2\delta\alpha_\delta^r + \delta^2\alpha_{\delta\delta}^r), \tag{40}$$

respectively. Here $\alpha_\delta^r = [\partial\alpha^r(\delta,\tau)/\partial\delta]_\tau$, $\alpha_{\delta\delta}^r = [\partial^2\alpha^r(\delta,\tau)/\partial\delta^2]_\tau$, $\alpha_{\delta\tau}^r = [\partial^2\alpha^r(\delta,\tau)/\partial\delta\partial\tau]$, $\alpha_\tau = [\partial\alpha(\delta,\tau)/\partial\tau]_\delta$, $\alpha_{\tau\tau} = [\partial^2\alpha(\delta,\tau)/\partial\tau^2]_\delta$, $\alpha^0(\delta,\tau)$ is the dimensionless Helmholtz energy of the ideal gas, $\alpha^r(\delta,\tau)$ is the residual part of the dimensionless Helmholtz energy, $\delta = \rho/\rho_c$ is the reduced density, $\tau = T_c/T$ is the inverse reduced temperature, $T_c = 150.687\ K$ is the critical temperature of argon, $\rho_c = 535.6\ kg\cdot m^{-3}$ is the critical density of argon, the values of the parameters $n_i, c_i, d_i, t_i, \eta_i, \varepsilon_i, \beta_i$ and $\gamma_i$ are given in Table 30 [17], and

$$\alpha(\delta,\tau) = \alpha^0(\delta,\tau) + \alpha^r(\delta,\tau), \tag{41}$$

$$\alpha^0(\delta,\tau) = \ln\delta + 8.31666243 - 4.94651164\tau + 1.5\cdot\ln\tau, \tag{42}$$

$$\alpha^r(\delta,\tau) = \sum_{i=1}^{12} n_i\delta^{d_i}\tau^{t_i} + \sum_{i=13}^{37} n_i\delta^{d_i}\tau^{t_i}\exp(-\delta^{c_i}) + \sum_{i=38}^{41} n_i\delta^{d_i}\tau^{t_i}\exp[-\eta_i(\delta-\varepsilon_i)^2 - \beta_i(\tau-\gamma_i)^2]. \tag{43}$$

The dimensionless Helmholtz energy $\alpha(\delta,\tau)$ (Eq. 41) is the Tegeler-Span-Wagner fundamental equation of state of argon (TSW-EOS) [17] which is used in the NIST database [14].

## 2.3. Comparison of predictions of the "meso-phase" hypothesis and TSW-EOS for argon

In order to compare the predictions of the "meso-phase" hypothesis with that of TSW-EOS we assume that $C_{VM}(\rho_B,T) = C_V(\rho_B,T)$, $A_{MB} = A(\rho_B,T)$, $S_{MB} = S(\rho_B,T)$, $E_{MB} = E(\rho_B,T)$, $G_{MB} = G(\rho_B,T)$ and $H_{MB} = H(\rho_B,T)$. We conclude from Eqs. 14-16 and 31-33 that $\chi_{TM}\rho kT/m = \Delta_M = s_{0M}$ and $\chi_T\rho kT/m = \Delta = s_0$, respectively. Eqs. 8, 16 and 27, 33 show that

$\omega_M m / kT = s_{0M}$ and $\omega m / kT = s_0$, respectively. Therefore we compare only the structural factors $s_{0M}$ and $s_0$. The reduced values are compared for the thermo-physical properties which have a dimensionality. We compare the isotherms corresponding to $T = T_{cM} = 151.2136\,K$, $T = 153\,K$, $155\,K$ and $157\,K$. The values of $\rho_A$ and $\rho_B$ at these temperatures are given in Table 1 [26].

The comparison of predictions of the "meso-phase" hypothesis and TSW-EOS are presented in Figs. 1, 2 and 3. The solid blue and dashed red lines correspond to the predictions of the reference TSW-EOS [17] and "meso-phase" hypothesis, respectively.

Fig. 1 presents the comparison of the isotherms of the relative pressure difference $(p_M / p - 1) \cdot 100\%$ (**a**, Eqs. 7 and 26), reduced isochoric heat capacity $C_V / k$ (**b**, Eqs. 4 and 28), reduced isobaric heat capacity $C_P / k$ (**c**, Eqs. 9 and 26) and reduced speed of sound $mc_s^2 / kT$ (**d**, Eqs. 10 and 33).

Fig. 1a shows that the relative pressure difference decreases with increasing temperature and it is much more than the inaccuracy of TSW-EOS. Fig. 1b shows that the difference between $C_V$ and $C_{VM}$ decreases with increasing temperature.

It was shown [30] that the expression for the isochoric heat capacity of liquid and gas, coexisting in phase equilibrium used in Woodcock's article [28] is incorrect. The continuous isochore of $C_V$ for $Ar$ was obtained in [28] by using the incorrect dependence of $C_V$ on temperature and density. The comparison of Fig. 1b [29] with Fig. 1b [28] shows that the isochores of argon in them are same.

The dependence of the isochoric heat capacity $C_V$ of argon along the isochore at the density value $13.3\,l^{-1}mol$ presented in Fig. 1b [29], which has no discontinuity, is incorrect because, according to experiments [4,15,16], $C_V$ along an isochore must have a discontinuity when the isochore of $C_V$ passes through the coexistence line.

It was shown above using Eq. 2 that the isochoric heat capacity in the "meso-phase" is equal to constant or it is a nonlinear function of density in the density interval $\rho_B \leq \rho \leq \rho_A$ in general case. As one can see from Fig. 1a [29] the isotherm of the isochoric heat capacity $C_{VM}$ for $T = 151\,K$ of argon in the "meso-phase" decreases linearly with increasing density. Hence, the isotherm of the isochoric heat capacity presented in Fig. 1a in [29] contradicts to Eq. 1, and the isotherm is incorrect if Eq. 1 correct and vice versa.

The density dependence of the isochoric heat capacity of argon obtained using the reference TSW-EOS [17] for $T = 151.2136\,K \approx 151\,K$ is presented in Fig. 1b. As on can see there is no density interval where the isochoric heat capacity decreases linearly with increasing density while according to the "meso-phase" hypothesis the density dependence of the isochoric heat capacity which is presented on Fig. 1a [29] decreases linearly with increasing density in the finite density interval.

Fig. 2 shows the comparison of the isotherms of the structural factor $s_0$ (**a**, Eqs. 16 and 33), reduced Helmholtz energy $A / kT$ (**b**, Eqs. 17 and 34), reduced Gibbs energy $G / kT$ (**c**, Eqs. 20 and 35) and reduced entropy $S / k$ (**d**, Eqs. 18 and 36).

Fig. 3 demonstrates the comparison of the isotherms of the reduced internal energy $E / kT$ (**a**, Eqs. 19 and 37), reduced enthalpy $H / kT$ (**b**, Eqs. 21 and 38), reduced Joule-Thompson coefficient $\mu k\rho / m$ (**c**, Eqs. 23 and 39) and the logarithm of the absolute value of the reduced isothermal throttling coefficient $|\delta_T|\rho/m$ (**d**, Eqs. 25 and 40). Note that $\delta_T$ has negative values at the isotherms.

One can see from Fig. 1c, 2a and 3d that $C_P$, $s_0$ and $\delta_T$ are finite at $T = T_{cM} = 151.2136\,K$ while $C_{PM}$, $s_{0M}$ and $\delta_T$ diverges at this temperature independently of the value of the density according to Eqs. 9, 16 and 25, respectively.

Figs. 2b, 2c, 2d, 3a, and 3b show that the "meso-phase" hypothesis is in the excellent agreement with the predictions of the TSW-EOS for the Helmholtz energy, Gibbs energy, entropy, internal energy and enthalpy of argon.

Figs. 1b, 1c, 1d, 2a, 3c and 3d show that the "meso-phase" hypothesis fails to describe quantitatively and qualitatively the isochoric and isobaric heat capacities, speed of sound, the structural factor $s_0$, Joule-Thompson coefficient and isothermal throttling coefficient of argon at the "meso-phase" region.

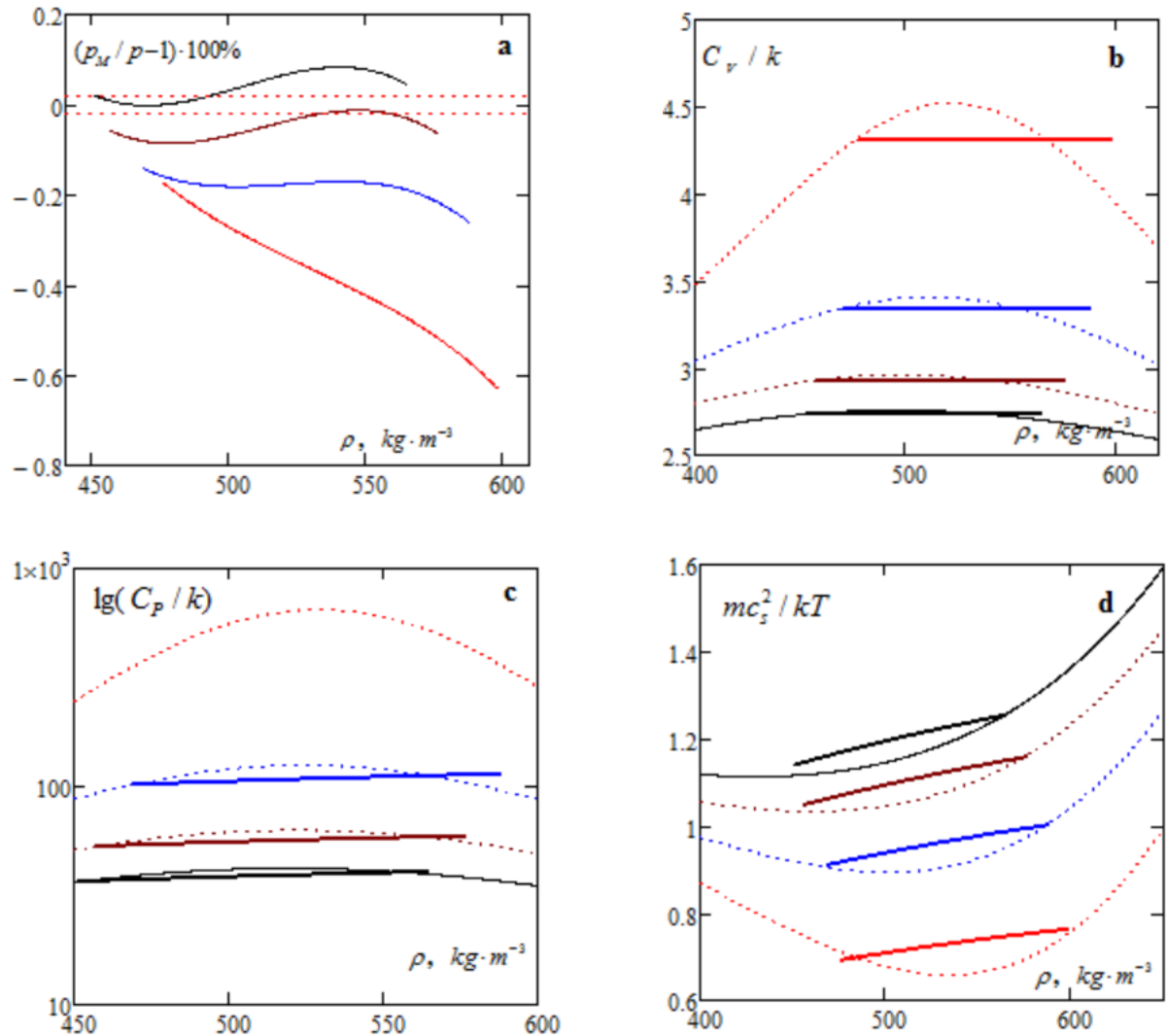

Fig. 1. The comparison of the isotherms of the relative pressure difference $(p_M/p-1)\cdot 100\%$ (Fig. 1**a**, Eqs. 7 and 26, the red dashed lines correspond to the inaccuracy $\pm 0.02\%$ of TSW-EOS, red solid line corresponds to $151.2136\,K$, blue solid line - $153\,K$, brown solid line - $155\,K$ and black solid line - $157\,K$), reduced isochoric heat capacities $C_V/k$ and $C_{VM}/k$ (Fig. 1**b**, Eqs. 4 and 28,),

decimal logarithm of the reduced isobaric heat capacities $C_P/k$ and $C_{PM}/k$ (Fig. 1c, Eqs. 9 and 26) and reduced speed of sound $mc_s^2/kT$ (Fig. 1d, Eqs. 10 and 33). The red dotted lines correspond to the predictions of TSW-EOS at $151.2136\,K$, blue dotted lines – $153\,K$, brown dotted lines - $155\,K$ and black solid lines - $157\,K$. The thick solid lines on Figs. 1b, 1c and 1d correspond to the predictions of the "meso-phase" hypothesis.

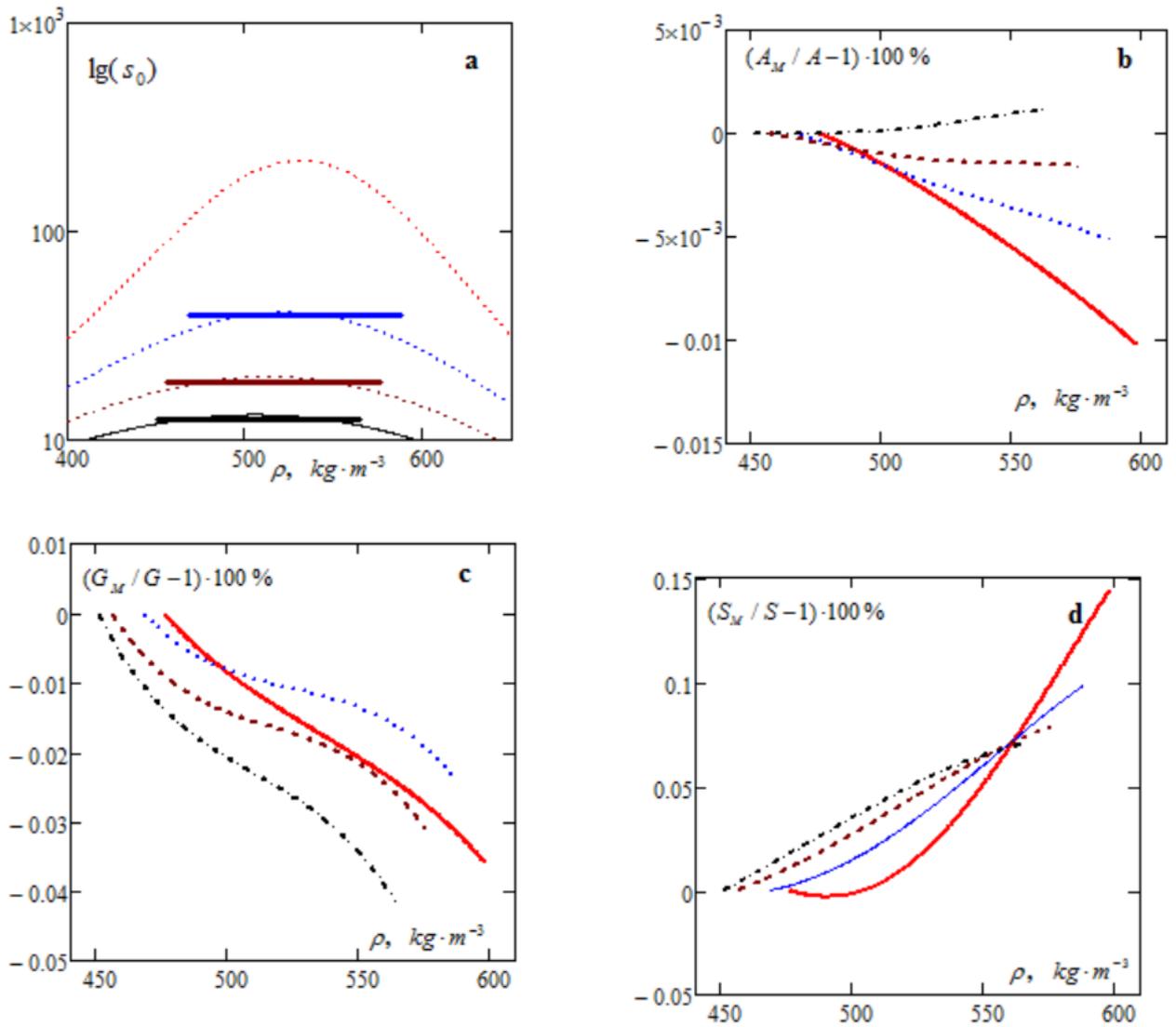

Fig. 2. The comparison of the isotherms of the decimal logarithm of the structural factor $s_0$ (a, Eqs. 16 and 33, the thick solid lines correspond to the predictions of the "meso-phase" hypothesis), relative Helmholtz energy difference $(A_M/A-1)\cdot 100\%$ (b, Eqs. 17 and 34), relative Gibbs energy difference $(G_M/G-1)\cdot 100\%$ (c, Eqs. 20 and 35) and relative entropy difference $(S_M/S-1)\cdot 100\%$ (d, Eqs. 18 and 36). The red solid lines correspond to $151.2136\,K$, blue solid lines - $153\,K$, brown solid lines - $155\,K$ and black solid lines - $157\,K$ on Figs. 2b, 2c and 2d.

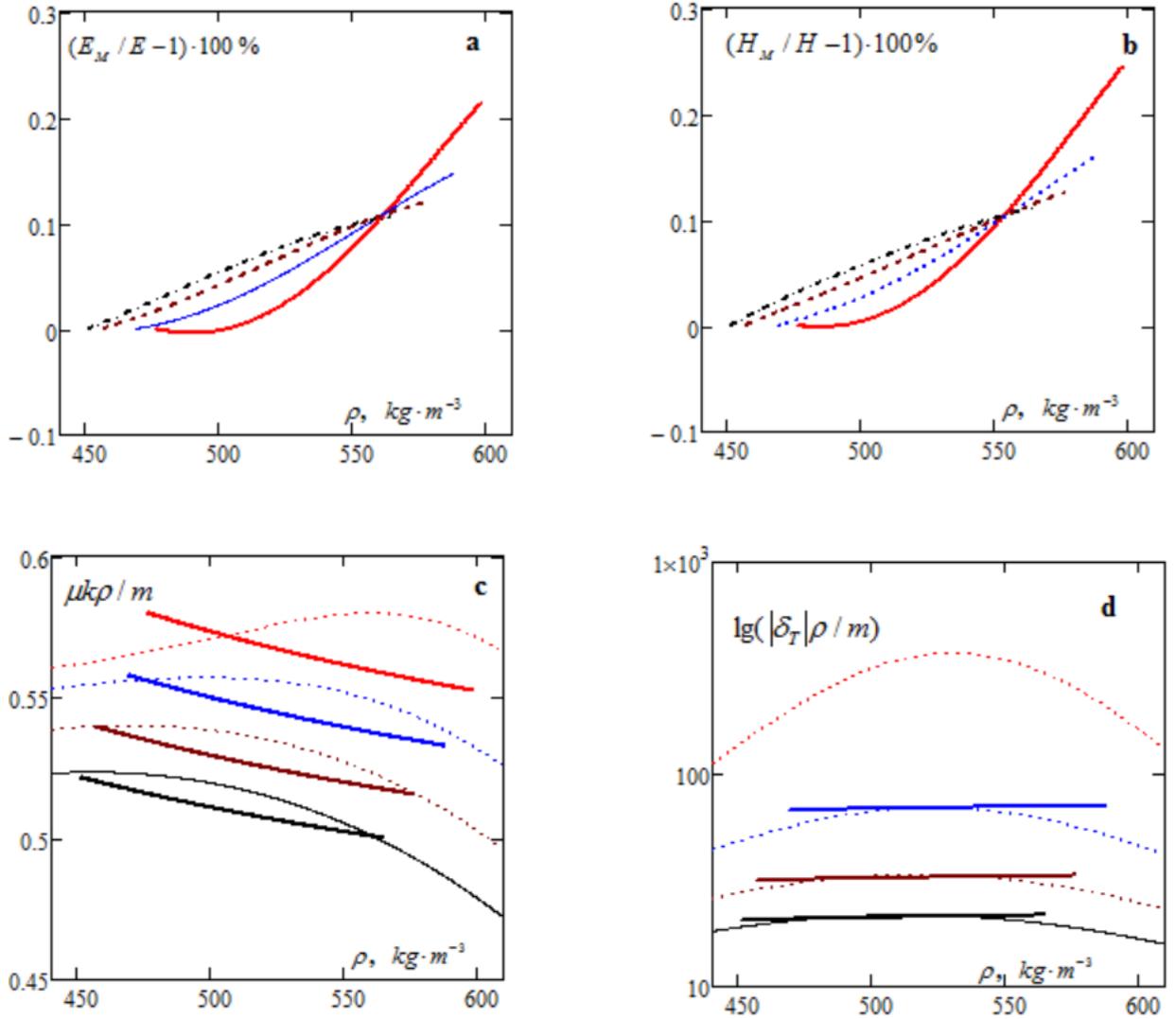

Fig. 3. The comparison of the isotherms of the reduced internal energy $(E_M/E-1)\cdot 100\%$ (**a**, Eqs. 19 and 37), reduced enthalpy $(H_M/H-1)\cdot 100\%$ (**b**, Eqs. 21 and 38), reduced Joule-Tompson coefficient $\mu k\rho/m$ (**c**, Eqs. 23 and 39) and decimal logarithm of the isothermal throttling coefficient $|\delta_T|\rho/m$ (**d**, Eqs. 25 and 40), corresponding to $T=T_{cM}$, $T=153\,K$, $155\,K$ and $157\,K$. The red solid lines correspond to $151.2136\,K$, blue solid lines - $153\,K$, brown solid lines - $155\,K$ and black solid lines - $157\,K$ on Figs. 3a and 3b. The thick solid lines on Figs. 3c and 3d correspond to the predictions of the "meso-phase" hypothesis.

## 3. About "failures" of VdW-EOS at the vapor–liquid critical region

According to [29] the liquid–gas critical point is not a property VdW-EOS can make any statements about, and VdW-EOS cannot describe qualitatively the excess Gibbs energy and isothermal rigidity coefficient of argon.

We show in Chapters **3.1** that VdW-EOS can describe qualitatively the excess Gibbs energy and isothermal rigidity coefficient of argon. Chapter **3.2** contains the discussion of the physically incorrect assertions of [29] concerning the temperature dependences of the isochoric heat capacity and entropy of the real fluids. Chapter **3.3** contains a response to the critique [29] of the parametric solution of the equations of the liquid-vapor coexistence of VdW-fluid. The quotes, statements, assertions and conclusions from [29] are italicized in Chapters **3.1**, **3.2**, **3.3** and **Appendix**.

### 3.1. The excess Gibbs energy and isothermal rigidity coefficient of argon

The reduced excess Gibbs energy $G_{red}$ is equal to $G_{red} \equiv (G^* - G_c^*)/kT$, where $G^* = G - G_{ig}$ is the excess Gibbs energy, $G$ is the Gibbs energy, $G_{ig}$ is the Gibbs energy of the ideal gas and $G_c^* = G^*\big|_{T_c, \rho_c}$. The critical isotherms of the reduced excess Gibbs energy for argon and VdW-fluid (the blue open circles) are presented in Fig. 4.

One can see using comparison Fig. 4 with Fig. 2b [29] that the critical isotherm of the reduced excess Gibbs energy for argon from Fig. 2b [29] is incorrect.

The isotherms of the reduced excess Gibbs energy for argon corresponding to $T_{cM}$ were defined from

$$G_{red}(\rho, T_{cM}) = [\alpha^r(\delta, \tau) + \delta\alpha_\delta^r(\delta, \tau)]\big|_{\tau = T_c/T_{cM}} - [\alpha^r(\delta, \tau) + \delta\alpha_\delta^r(\delta, \tau)]\big|_{\delta=1, \tau = T_c/T_{cM}}. \tag{44}$$

Eq. 44 is obtained using the relation $G_{ig}(\rho,T)/kT = 1 + \alpha^0(\delta, \tau)$ and Eqs. 35 and 41.

The reduced excess Gibbs energy of VdW-fluid is defined from the relation

$$G_{red} = [G^*(V,T) - G^*(V_c, T_c)]/RT, \tag{45}$$

where the excess Gibbs energy $G^*(V,T)$ of VdW-fluid

$$G^*(V,T) = bRT/(V-b) - RT\ln(1 - b/V) - 2a/V \tag{46}$$

was obtained using the relations $p = -(\partial A/\partial V)_T$ and $G = A + pV$ from the Van der Waals' equation of state [6]

$$p(n,T,a,b) = nkT/(1-bn) - an^2, \tag{47}$$

where $n = 1/v$ is the number density, and $a$ and $b$ are positive constants.

Eq. 47 can be presented as the reduced equation of state of VdW-fluid

$$\bar{p}_r(\bar{n}_r, \bar{T}_r) = 8\bar{n}_r\bar{T}_r/(3 - \bar{n}_r) - 3\bar{n}_r^2, \tag{48}$$

where $\bar{p}_r = p/\bar{p}_c$, $\bar{T}_r = T/\bar{T}_c$ and $\bar{n}_r = \bar{v}_c/v = n/\bar{n}_c$ are the reduced pressure, temperature and density of VdW-fluid, respectively, and $\bar{p}_c$, $\bar{T}_c$, $\bar{v}_c$, $\bar{n}_c = 1/\bar{v}_c$ and $\bar{z}_c = \bar{p}_c/\bar{n}_ck\bar{T}_c$ are the critical pressure, temperature, volume, number density and compressibility factor of VdW-fluid, respectively. The values $\bar{p}_c = a/27b^2$, $\bar{T}_c = 8/27kb$ and $\bar{n}_c = 1/\bar{v}_c = 1/3b$ are defined using the following conditions of the critical point

$$\bar{p}_c = p(\bar{n}_c, \bar{T}_c, a, b),\ \partial p(n,T,a,b)/\partial n\big|_{n=\bar{n}_c, T=\bar{T}_c} = 0,\ \partial^2 p(n,T,a,b)/\partial n^2\big|_{n=\bar{n}_c, T=\bar{T}_c} = 0. \tag{49}$$

We obtain from Eq. 47 the reduced VdW-EOS

$$p_r(n_r, T_r, z_c, p_c, n_c, a, b) = n_rT_r/z_c(1 - n_rbn_c) - n_r^2an_c^2/p_c, \tag{50}$$

where $p_r = p/p_c$, $T_r = T/T_c$ and $n_r = v_c/v = n/n_c$ are the reduced pressure, temperature and density of the real fluid, respectively, and $p_c$, $T_c$, $v_c$, $n_c = 1/v_c$ and $z_c = p_c v_c / kT_c$ are the critical pressure, temperature, molar volume, number density and compressibility factor of the real fluid, respectively.

According to the corresponding states principle [8-10] one can replace $\bar{p}_r$, $\bar{T}_r$ and $\bar{n}_r$ for VdW-fluid in Eq. 48 by $p_r$, $T_r$ and $n_r$ of the real fluid, respectively, to obtain new equation of state

$$p_r(n_r, T_r) = 8n_r T_r /(3-n_r) - 3n_r^2. \tag{51}$$

It is easy to see that VdW-EOS defines the exact position of the critical point on the thermodynamic (temperature, pressure)- , (density, pressure)- and (density, temperature)- planes if the coefficients $a$ and $b$ of VdW-EOS are defined from ($p_c, T_c$), ($p_c, v_c$) and ($T_c, v_c$) using the relations

$$a_1 = 27k^2 T_c^2 / 64 p_c, \quad b_1 = kT_c / 8 p_c, \tag{52}$$
$$a_2 = 3 p_c v_c^2, \quad b_2 = v_c / 3, \tag{53}$$
$$a_3 = 9kT_c v_c / 8, \quad b_3 = v_c / 3, \tag{54}$$

respectively. We obtain for argon having $p_c = 4.863\, MPa$ [17,21,22]

$$a_1 = 1.337\, l^2 mol^{-2} atm, \quad b_1 = 0.302\, mol^{-1} l, \tag{55}$$
$$a_2 = 2.43\, l^2 mol^{-2} atm, \quad b_2 = 0.025\, mol^{-1} l, \tag{56}$$
$$a_3 = 1.045\, l^2 mol^{-2} atm, \quad b_3 = 0.025\, mol^{-1} l. \tag{57}$$

We obtain the relations

$$G_{red}(\rho, T_{cM}) = \delta/(8z_c - \delta) - 1/(8z_c - 1) + \ln[(8z_c - 1)/(8z_c - \delta)] + 27(1 - T_c \delta / T_{cM})/32 z_c, \tag{58}$$
$$G_{red}(\rho, T_{cM}) = \delta/(3-\delta) - 1/2 + \ln[2/(3-\delta)] + 6z_c(1 - T_c \delta / T_{cM}), \tag{59}$$
$$G_{red}(\rho, T_{cM}) = \delta/(3-\delta) - 1/2 + \ln[2/(3-\delta)] + 9(1 - T_c \delta / T_{cM})/4, \tag{60}$$

from Eqs. 48, 50 and 51, respectively. The critical isotherms of the reduced excess Gibbs energy $G_{red}$ for VdW-fluid was calculated using Eqs. 58-60.

The comparison of Fig. 4 with Fig. 2a [29] shows that VdW-EOS describes the excess Gibbs energy of argon in the critical region. Therefore, the comparison of the dependencies presented in Figs. 2a and 2b [29] is incorrect, and the statements "*Gibbs energy of argon, taken from the NIST thermophysical property tables [7], by comparison shows that the van der Waals equation completely misses the essential behavior, especially in the vicinity of the critical point*", "*the absurd minimum $G^*$ at $\rho \sim 20\, mol \cdot l^{-1}$ and subsequent increase for the hypothetical van der Waals liquid are consequences of $V < b$ in Eq. 1 at this density*", and "*it is evident from Fig. 2a, b that the van der Waals equation fails to describe even qualitatively the thermodynamic properties of gas–liquid coexistence in the critical region*" [29] are incorrect.

Fig. 4 shows that the minimum of $G^*$ at $\rho \sim 20\, l^{-1} mol$ for VdW-fluid is not absurd and VdW-EOS can describe quantitatively the excess Gibbs energy of argon in the critical region.

As one can see from Fig. 2a [29] the value $V \sim 0.05\ mol^{-1}l$ corresponds to $\rho \sim 20\ l^{-1}mol$. So, $V > b = 0.0320\ mol^{-1}l$. Therefore, the assertion "*The absurd minimum $G^*$ at $\rho \sim 20\ mol\cdot l^{-1}$ and subsequent increase for the hypothetical van der Waals liquid are consequences of $V < b$ in Eq. 1 at this density*" [29] is incorrect.

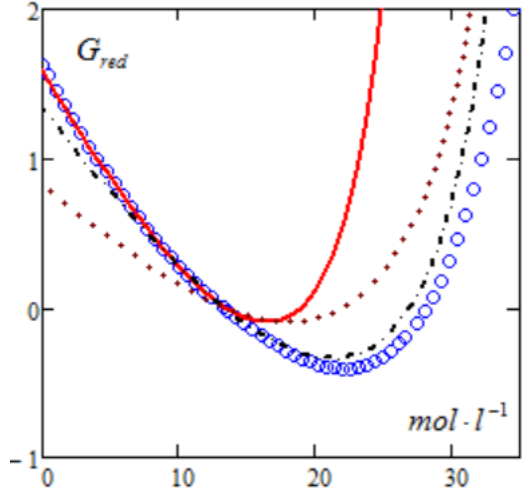

Fig. 4. The critical isotherms of the reduced excess Gibbs energy of argon (Eq. 44) (blue open circles) and VdW-fluid: the solid red line corresponds to Eq. 58, the dotted brown line corresponds to Eq. 59 and the dashed-dotted black line corresponds to Eq. 60.

We have from Eq. 47 for the reduced isothermal rigidity coefficient

$$\omega m/kT = 1/(1-b/V)^2 - 2a/RTV. \tag{61}$$

We have from Eq. 61 the relations

$$\omega(\rho,T)m/kT = (1-\delta/8z_c)^{-2} - 27\delta\tau/32z_c, \tag{62}$$

$$\omega(\rho,T)m/kT = (1-\delta/3)^{-2} - 6z_c\delta\tau, \tag{63}$$

$$\omega(\rho,T)m/kT = (1-\delta/3)^{-2} - 9\delta\tau/4, \tag{64}$$

corresponding to Eqs. 48, 50 and 51, respectively.

Fig. 5 presents the comparison of the reduced isothermal rigidity coefficient for argon (Eq. 27, blue open circles) with the predictions of VdW-EOS along critical isotherm in the vicinity of the critical point. The critical isotherms $\omega(\rho,T_c)m/kT_c$ of the reduced isothermal rigidity coefficient for VdW-fluid was calculated using Eqs. 62-64. One can see from Fig. 5 that VdW-EOS can describe qualitatively the critical isotherm of the reduced isothermal rigidity coefficient of argon which is defined from Eq. 27.

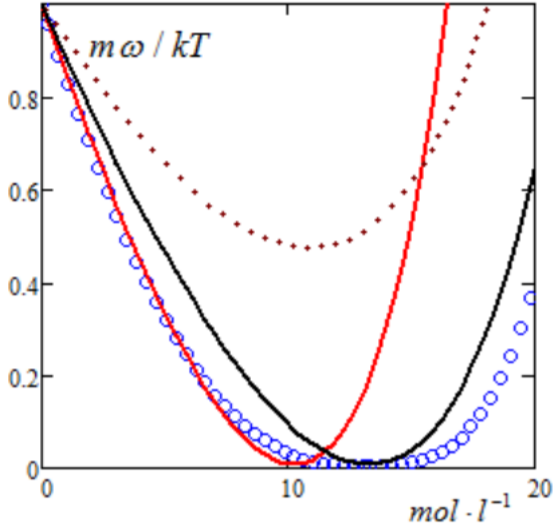

Fig. 5. The reduced isothermal rigidity coefficient $m\omega/kT$ for argon (Eq. 27, blue open circles) compared with the prediction of VdW-EOS along the isotherm $T_{cM} = 151.2136\,K$ in the vicinity of the critical point. Solid red line corresponds to Eq. 62, the dotted brown line corresponds to Eq. 63, solid black line corresponds to Eq. 64.

It is easy to establish from Eqs. 1-6 [29] that Eqs. 2-6 [29] for excess state functions relative to an ideal gas $(V \to \infty)$ are incorrect and they must be replaced by

$$A^* = -\int_\infty^V P^* dV = -RT\ln[(V-b)/V] - a/V,$$

$$S^* = \int_\infty^V (\partial P^*/\partial T)_V dV = R\ln[(V-b)/V],$$

$$U^* = A^* + TS^* = -a/V,$$

$$H^* = U^* + P^*V = RTb/(V-b) - 2a/V,$$

$$G^* = H^* - TS^* = RT[b/(V-b) - \ln(1-b/V)] - 2a/V,$$

where $P^*(T,V) \equiv P(T,V) - RT/V = RTb/V(V-b) - a/V^2$, which is obtained from Eq. 47.

The definitions $\omega \equiv (\partial p/\partial \rho)_T$ and $\rho \equiv 1/V$ were used in [29]. We obtain $\omega = -V^2(\partial p/\partial V)_T$ and $(\partial \omega/\partial \rho)_T = V^2 (\partial [V^2(\partial p/\partial V)_T]/\partial V)_T = -2V\omega + V^4(\partial^2 p/\partial V^2)_T$. Therefore we can conclude that Eq. 9 [29] is incorrect and it must be replaced by the following correct equation

$$"H"(y \text{ in ref. 4}) = b^4/a \cdot (\partial^2 p/\partial V^2)_T = b^4/a \cdot [2\omega/V^3 + (\partial \omega/\partial \rho)_T/V^4].$$

**3.2. VdW-EOS and the isochoric heat capacity and entropy of the ideal gas**

VdW-EOS predicts that $C_V$ is equal to that of the ideal gas $C_{V,ig}$ [7,8,23-25]. According to [8] $C_{V,ig}$ per particle (atom or molecule) in general case is a function of temperature and it does not depend on density; it is equal to $3k/2$ for atomic substances at temperatures $T \gg \hbar^2/kmv^{2/3}$, where $\hbar$ is the Planck's constant, $m$ is an atomic mass; $C_{V,ig}$ is equal to constant, which differs from $3k/2$, for the rigid rotator model of molecule at $T \gg \max\{\hbar^2/kMv^{2/3}, \hbar^2/kI_{\min}\}$, where $M$ is a molecular mass, $I_{\min} = \min\{I_1, I_2, I_3\}$, $I_1$, $I_2$ and $I_3$ are the principal momenta of inertia of a nonlinear molecule, and $I_{\min} = I$, where $I$ is a moment of inertia of a linear molecule; $C_{V,ig}$ is equal to constant, which is greater than $3k/2$, for molecule consisting of $n$ atoms with masses $m_i$, $i=1,..,n$, at

$T >> \max\{\hbar^2/kMv^{2/3}, \hbar^2/kI_{\min}, \hbar\omega_{\max}/k\}$, where $M = \sum_{i=1}^{n} m_i$, $\omega_{\max} = \max\{\omega_j, i=1,...,3n-6\}$, $\omega_j$ is a frequency of $j$-th harmonic vibration mode of a nonlinear molecule, and $\omega_{\max} = \max\{\omega_{lin,j}, i=1, ..., 3n-5\}$, where $\omega_{lin,j}$ is a frequency of $j$-th harmonic vibration mode of a linear molecule; $C_{V,ig}$ of the atomic fluids differs from that of molecular fluids; $C_{V,ig}$ of molecular fluids depends, particularly, on the spatial structure and masses of the atoms consisting the molecule as well as interactions between the atoms; and $C_{V,ig}$ of various molecular fluids can differ from each other. So, the statements "*Van der Waals' equation ... erroneously predicts, for instance, that $C_V$ is a constant for all fluid states*", "*van der Waals equation predicts the same heat capacity $(3R/2)$ for all thermodynamic states of all fluids*", and "*Equation 1 ... predicts that all fluids have a constant $C_V$, i.e. equal to that of the ideal gas $(3R/2)$*" [29] are incorrect.

The entropy (per molecule) of the ideal gas $S_{ig}$ consisting of molecules depends on the temperature. For example, entropy of the molecule consisting of two different atoms with masses $m_1$ and $m_2$ which is approximately equal to [8]

$$S_{ig} = \left(\frac{\partial}{\partial T}\left[kT\ln\left(ev\left(\frac{MkT}{2\pi\hbar^2}\right)^{3/2}\sum_{i=0}^{\infty}\exp\left(-\frac{\hbar\Omega}{kT}(i+1/2)\right)\cdot\sum_{j=0}^{\infty}(2j+1)\exp\left(\frac{-\hbar^2 j(j+1)}{2IkT}\right)\right)\right]\right)_v,$$

depends on temperature. Therefore, the statement "*Entropy of the ideal gas is independent of temperature at constant volume*" [29] is incorrect. Here $M = m_1 + m_2$ is the mass of the molecule, $I$ is the moment of the inertia of the molecule, $\Omega$ is the frequency of linear (harmonic) oscillations of the molecule.

The entropy of the ideal gas $S_{ig}$ depends on temperature in general case. So, it is clear that the equality $\Delta S_{ig} = Q_{rev}/T$, were $\Delta S_{ig}$ is the change of entropy of the ideal gas, may be valid if the heat $Q_{rev}$ is added reversibly to a real fluid at constant volume $V$. Therefore, the assertion "*by definition, $\Delta S = Q_{rev}/T$ (where $Q_{rev}$ is reversible heat added), $S^*$ must increase to some extent with $T$ if heat is added reversibly to a real fluid at constant $V$*" [29] could be incorrect.

It was shown earlier in [13] that VdW-EOS near critical point can be presented in an asymptotic form of the equation of state of scaling theory. So, the assertion "*Van der Waals equation, however, is inconsistent with the universal scaling singularity concept*" [29] is incorrect.

### 3.3. Properties of parametric solution of equations of liquid-gas coexistence of VdW-fluid

According to the parametric solution [27] of the equations corresponding to the liquid-vapor phase equilibrium of VdW-fluid

$$bkT/a = F(y(T)) \equiv \left.\frac{2(y - e^{2y} + ye^{2y} + 1)(4ye^{2y} - e^{4y} + 1)^2}{(e^{2y} - 1)(2y - 2e^{2y} + e^{4y} - 2ye^{4y} + 4y^2e^{2y} + 1)^2}\right|_{y=y(T)}, \qquad (65)$$

$$bn_L(T) = F_L(y(T)) = \left.2\cdot\frac{y - e^{2y} + ye^{2y} + 1}{(2y-1)e^{2y} - e^{-2y} - 2y + 2}\right|_{y=y(T)}, \qquad (66)$$

$$bn_V(T) = F_V(y(T)) \equiv 2 \cdot \left. \frac{y - e^{2y} + ye^{2y} + 1}{e^{4y} - e^{2y}(2y+2) + 2y + 1} \right|_{y=y(T)}, \tag{67}$$

$$p_e(T) = a/b^2 \cdot [FF_L/(1-F_L) - F_L^2]\big|_{y=y(T)}, \tag{68}$$

where the temperature dependence of the parameter $y(T)$ is defined from Eq. 65. The temperature dependencies of the saturation pressure $p_e(T)$ and the densities of liquid $n_L(T) = 1/v_L(T)$ and vapor $n_V(T) = 1/v_V(T)$ of VdW-fluid are defined from Eqs. 66-68.

One can obtain from Eq. 47 using the relations $p = -(\partial A/\partial v)_T$ and $S = -(\partial A/\partial T)_V$ the relation $(S_V - S_L)/k = \ln[(1/bn_V - 1)/(1/bn_L - 1)]$ for the difference of the entropies of vapor $S_V$ and liquid $S_L$ coexisting in phase equilibrium. According to [27] $2y = \ln[(1/bn_V - 1)/(1/bn_L - 1)]$. Hence we conclude that $y = (S_V - S_L)/2k$. So the parameter $y$ is equal to the half of the reduced coexistence entropy difference [23-25]. We obtain $q = 2a/b \cdot yF(y)$ for the latent heat of vaporization because $q = T(S_V - S_L)$ [8]. So the parameter $y$ is not a formal parameter, and it has the physical sense.

A correct comparison of the phase equilibrium line of VdW-fluid with that of real fluid implies the definition of $n_L(T)$ and $n_V(T)$ from the above Eqs. 66-67.

The temperature dependence of the parameter $y$ is defined by Woodcock [29] from $y_W(T) = 0.5 \cdot \ln([mN_A/b\rho_{gas}(T) - 1]/[mN_A/b\rho_{liq}(T) - 1])$, where $N_A$ is the Avogadro number, $\rho_{gas}$ and $\rho_{liq}$ are the mass densities of the liquid and vapor of the real fluid (argon) coexisting in phase equilibrium; then he defines some functions $\rho_{L,W}(T)$ and $\rho_{G,W}(T)$ from $b\rho_{L,W}(T)/mN_A = F_L(y_W(T))$ and $b\rho_{V,W}(T)/mN_A = F_V(y_W(T))$. It is clear that: $y_W(T)$ is not the parameter of VdW-fluid, so, $y_W(T) \neq y(T)$; $\rho_{L,W}$ and $\rho_{G,W}$ are not the densities of the liquid and vapor of VdW-fluid; and $\rho_{L,W} \neq mN_A n_L$ and $\rho_{V,W} \neq mN_A n_V$ because $y(T) \neq y_W(T)$. It is easy to see that: the values of $y_W(T)$ at critical temperature are presented in the last column of Table 1 [29]; the dependence $y_W(T)$ is presented in Fig. 3 [29]; the functions $F_L(y_W(T))$ and $F_V(y_W(T))$ are presented by the solid blue lines in fig. 4 [29]; the rigidity $\omega$ which is defined from Eq. 8 [29] using $y_W(T)$ is presented by the solid blue line in Fig. 5 [29]. So, the comparisons made by using the last column of Table 1 [29] and Figs. 3-5 [29] do not concern VdW-fluid.

So, we have shown that the dependencies presented in [29] for the coexisting difference functional of argon and coexisting densities of liquid and vapor of VdW-fluid are incorrect, and Table 1 [29] includes incorrect values of coexisting difference functional. It is clear that the comparisons and the conclusions in [29] based on $y_W(T)$ have no sense.

As one can see from Eqs. 4-5 [27], Fig. 1 [27] and Fig. 6, the difference between the densities of liquid and gas coexisting in the phase equilibrium vanishes when $y = 0$. So, the statements *"The coexistence density difference function of $y$, $\Delta F(y) = b(\rho_{liq} - \rho_{gas})_{coex}$ must go to zero when $y = 1$. Plotting $F_{gas}(y)$ and $F_{liq}(y)$ against $y$, and finding that they have the singular value $\rho_c b = 1/3$ when $y = 1$ does not prove anything; there is no basis for assertion 2 above"* [29] are incorrect.

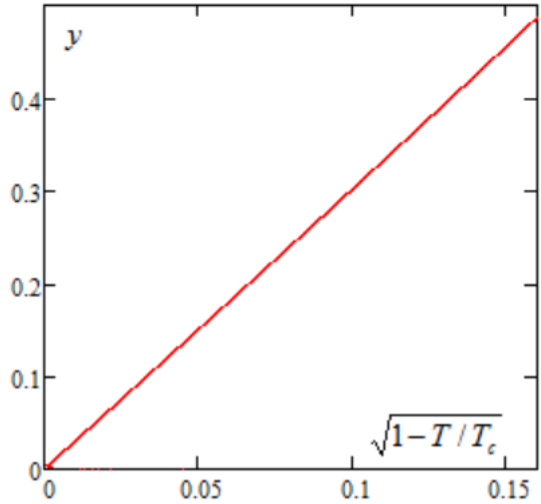

Fig. 6. The dependence of the parameter $y$ (the coexistence density functional [29]) of VdW-fluid on $\sqrt{1-T/T_c}$ which is obtained from Eq. 59.

According to Fig. 2 [27] the first and second partial derivatives of pressure with respect to volume at constant temperature go to zero in the limit $y \to 0+$, which means that $T_c$ is reached from the side of low temperatures (see Eq. 6 and Fig. 1 [27]). According to [29], the rigidity $\omega$ and its density derivatives go to zero for real gas and liquid states at $T_c$, $p_c$, if $T_c$ is reached from the side of high temperatures. Therefore, the conclusion "*Figure 2 in Ref. [4], showing that these two derivatives go to zero when $y = 1$, does not prove anything because $\omega$ and its density derivatives all go to zero for real gas and liquid states at $T_c$, $p_c$. This is illustrated in Fig. 5 for the behavior of the rigidity of argon along the critical isotherm, compared to the prediction of van der Waals equation*" [29] are incorrect.

One can conclude using Eq. 6 [27] that the inequalities $0 \le y \le 0.5$ which are valid for Fig. 2 [27] correspond to the temperature interval $147\,K \le T \le 151\,K$ for $a = 1.337\,l^2 mol^{-2} atm$ and $b = 0.0320\,mol^{-1}l$ used in [29]. Therefore, the assertion "*the rigidity is related to the two reduced derivatives introduced in Ref. [4] and plotted against y for a very narrow $(<1\,K)$ near-critical range in figure 2 of [4]*" [29] is incorrect.

The parameter $y$ was used earlier in [23-25] which were cited in [29]. As one can see from the definition of the parameter $y = 0.5\ln[(1/bn_V - 1)/(1/bn_L - 1)]$ it depends on the constant $b$ and saturation densities of the liquid $n_L$ and vapor $n_V$ of VdW-fluid. $n_L$ and $n_V$ are defined from the thermodynamic conditions of the phase equilibrium of VdW-fluid which are defined by VdW-EOS (see Eqs. 1-3 [27]). Hence, $y$ depends on the constant $a$ of VdW-EOS too. So, the parameter $y$ is not defined independently of VdW-EOS functionally. Therefore, the conclusion "*coexistence state function $y(T)$ is defined independently of van der Waals equation functionally*" [29] is incorrect.

The critical temperature $T_c$ must have a positive value [1-25]. So, the statement "*The density difference, $\Delta F(y) = F_{liq}(y) - F_{gas}(y)$ (see figure 1 of Ref. [4]) does not go to zero $T_c$ in the case of a real fluid*" [29] has no sense.

One can see from Figs. 1 and 2 [27] that the functions $\Delta F = F_L - F_V$, $H_L$, $H_V$, $G_L$ and $G_V$ vanish at $y = 0$, therefore, the statements "*the fact that "$\Delta F$", "$H$" and "$G$" go to zero at

$y = 1$ *for both coexisting gas and liquid in figures 1 and 2 of Ref. [4], respectively, does not prove anything about criticality of real fluids*" [29] have no sense.

According to the scaling theory which has a strong physical basis and quantitatively describes the thermodynamic properties of fluid near critical point [1,8], the density difference between gas and liquid vanishes at critical point and the temperature dependencies of saturation densities of the gas and liquid near critical point are determined by the equations $\rho_{liq}/\rho_c = 1 + c(T_c - T)^\beta$, $\rho_{gas}/\rho_c = 1 - c(T_c - T)^\beta$, where $c > 0$, $\beta > 0$ and $\beta \neq 1/2$. One can conclude using Eq. 48 that the parameter $y$ vanishes at critical temperature. Therefore the conclusion "$y(T)$ *interpolates to a constant nonzero value at $T_c$, and there is no evidence, experimental or otherwise, nor any good theoretical reason to believe any departure from this result within a tiny fraction of 1 degree K below $T_c$*" [29] is not correct.

## 4. Conclusion

We showed that the "meso-phase" hypothesis of Woodcock L. V. fails to describe quantitatively and qualitatively the isochoric and isobaric heat capacities, speed of sound, long wavelength limit of the structural factor, isothermal compressibility, density fluctuations, Joule-Thompson coefficient and isothermal throttling coefficient of argon in the "meso-phase" region. It is also shown that VdW-EOS can describe qualitatively the excess Gibbs energy and rigidity of argon near critical point.

It is shown that: the dependencies for the isochoric heat capacity, excess Gibbs energy and coexisting difference functional of argon, and coexisting densities of liquid and vapor of VdW-fluid presented in all Figures in the paper [29] are incorrect; Table 1 [29] includes incorrect values of coexisting difference functional; [29] includes many incorrect equations, mathematical and logical errors, incorrect comparisons and incorrect assertions concerning the temperature dependences of the isochoric heat capacity and entropy of the real fluids; most of the conclusions in [29] are based on the above errors, incorrect data, incorrect comparisons and incorrect dependences. Therefore, the most of conclusions in [29] are not valid.

# Appendix

Let us consider the first assertion "*In contrast to the conjecture [1] there is no reliable experimental evidence to doubt the existence of a single critical point,*" citing the Sengers and Anisimov comment [2] based upon historic evidence from divergent isochoric heat capacity $C_V$ measurements at the critical temperature ($T_c$)" which was discussed in [29].

The first part of the assertion ("*In contrast to the conjecture [1] there is no reliable experimental evidence to doubt the existence of a single critical point*") was quoted from [4] in [27], but there was not the rest part of the assertion ("*citing the Sengers and Anisimov comment [2] based upon historic evidence from divergent isochoric heat capacity $C_V$ measurements at the critical temperature ($T_c$)*") in [27]. The first part of the assertion means that there is no reliable experimental evidence to doubt the existence of a single critical point and this is in contrast to the conjecture of [29] and nothing more. So, the first assertion discussed in [29] is an incorrect assertion from [27], while a correct assertion from [27] is:

Assertion 1. "In contrast to the conjecture [1] there is no reliable experimental evidence to doubt the existence of a single critical point".

From logical point of view, it is clear that an experimental proof of the existence of two or more critical points or the existence of a critical line will be the proof of the incorrectness of the Assertion 1. However, such experimental proof was not presented in [29]. Moreover, one can see from [29] that there are no other proofs in [29] for the Assertion 1 to be incorrect.

It is evident that the Assertion 1 does not mean that Anisimov and Sengers divergent $C_V$ at $T_c$ is wrong. Therefore, the conclusions "*if Umirzakov's first assertion were to be right, Anisimov and Sengers divergent $C_V$ at $T_c$ would have to be wrong. In fact, neither of the assertions will withstand scientific scrutiny*" [29] have no sense.

The second assertion discussed in [29] is "*… to prove that the existence of a single critical point of a fluid described by van der Waals equation of state (VDW-EOS) is not a hypothesis and is a consequence of the thermodynamic conditions of liquid–vapor phase equilibrium.*"

One can see from [27] that the quote in the second assertion is incorrect and a correct assertion from [27] is:

Assertion 2. "We prove that the existence of a single critical point of a fluid described by van der Waals equation of state (VDW-EOS) is not a hypothesis and is a consequence of the thermodynamic conditions of liquid–vapor phase equilibrium".

It is easy to see reading [29] that there is no proof in [29] that the existence of a single critical point of the fluid described by VdW-EOS is hypothetical and the existence of a single critical point of VdW-fluid is not a consequence of the thermodynamic conditions of liquid–vapor phase equilibrium. So, there is no proof in [29] that the Assertion 2 is incorrect.

One can see that VdW-EOS [6] alone was considered in [27] and all conclusions of [27] concern VdW-fluid. There is no statement or assumption in [27] that VDW-EOS describes quantitatively the thermodynamic properties of the real fluids. It is evident that the statements of [27] that "there is no reliable experimental evidence to doubt the existence of a single critical point" and "the existence of a single critical point of a fluid described by the van der Waals equation of state (VDW-EOS) is not a hypothesis and is a consequence of the thermodynamic conditions of liquid–vapor phase equilibrium" do not mean that VdW-EOS describes quantitatively the thermodynamic properties of the real fluids (for example, of argon). It is also evident that the proof that VdW-EOS cannot describe quantitatively the thermodynamic properties of the real fluids does not mean that the above statements of [27] are incorrect.

According to [29], "*it was incorrectly asserted that van der Waals equation "proves" the existence of a scaling singularity with a divergent isochoric heat capacity ( $C_V$ )*" in [27]. One can easily see from [27] that there is no assertion in [27] that VdW-EOS proves the existence of a scaling singularity with a divergent isochoric heat capacity.

One can see from the above comments that there is the lack of logic in the reasoning of Woodcock in [29].

We proved in [27] that VdW-fluid has only one critical point. Therefore, the statement "*Ref. [1] proves nothing more than van der Waals' equation has a singularity with two vanishing derivatives*" [29] is incorrect if the singularity does not mean that there is only one critical point.

The ability of VdW-EOS to describe the thermodynamic properties of real fluid was not considered in [27]. The fact that VdW-EOS cannot describe quantitatively the thermodynamic properties of the real fluids was earlier established by many authors [7-9,23]. So, the statement in [29] that "*state functions of van der Waal's equation fail to describe the thermodynamic properties of low-temperature gases, liquids and gas-liquid coexistence*" is not a new insight into the science or physics.

Many conclusions in [29] are based on the fact that VdW-EOS cannot describe quantitatively the thermodynamic properties of the real fluids. This fact does not prove the statements such as "*The conclusion that there is no "critical point" singularity on Gibbs density surface remains scientifically sound*", "*the conclusion in Ref. [1], i.e., that there is no critical point singularity with scaling properties on Gibbs density surface still holds true*", and "*Van der Waals hypothetical singular critical point is based upon a common misconception that van der Waals equation represents physical reality of fluids*" [29].

According to [29] "*Explicitly built into the equation is an incorrect a priori assumption of continuity of liquid and gaseous states*". One can see from the detailed consideration of [29] that there is no proof of the incorrectness of a priori assumption of continuity of liquid and gaseous states in [29].

There exists the method for direct experimental measure of a critical density – the disappearance of the meniscus method which gives a high precision of the critical density determination (±0.02%) [12,31-33]. The radioactive tracer technique is also used for direct measurement of the critical density [34]. So, the statement "*No research in history has reported the direct experimental measurement of a critical density*" [29] is incorrect.

One can see from comparison of contents of [27] and [29] that [29] does not include the proofs of the incorrectness of the assertions and conclusions made in [27]. One can also see the same from the comments presented above.